\documentclass[aps, prb, reprint, amsmath, amssymb, superscriptaddress, showpacs, floatfix]{revtex4-2}

\usepackage{mathtools}
\usepackage{bm}
\usepackage{textgreek}
\usepackage{mleftright}
\mleftright

\usepackage[dvipsnames,table,hyperref,rgb]{xcolor}
\definecolor{linkcolor}{rgb}{0.16, 0.32, 0.75}
\definecolor{citecolor}{rgb}{0.0, 0.5, 0.0}
\definecolor{urlcolor}{rgb}{0.06, 0.46, 1.0}
\usepackage[colorlinks, linkcolor={linkcolor}, citecolor={citecolor}, urlcolor={urlcolor}]{hyperref}

\begin{document}

\title{Andreev conductance in disordered SF junctions with spin-orbit scattering}

\author{M.\,E.\,Ismagambetov}

\affiliation{Max Planck Institute for Solid State Research, 70569 Stuttgart, Germany}

\author{P.\,M.\,Ostrovsky}

\affiliation{TKM, Karlsruhe Institute of Technology, 76131 Karlsruhe, Germany}
\affiliation{Max Planck Institute for Solid State Research, 70569 Stuttgart, Germany}

\author{M.\,V.\,Feigel'man}

\affiliation{Nanocenter CENN, Jamova 39, SI-1000 Ljubljana, Slovenia}
\affiliation{L.\,D.\,Landau Institute for Theoretical Physics, Chernogolovka 142432, Russia}

\begin{abstract}
We calculate the conductance of a junction between a disordered superconductor and a very strong half-metallic ferromagnet admitting electrons with only one spin projection. A usual mechanism of Andreev reflection is strongly suppressed in this case since Cooper pairs are composed of electrons with opposite spins. However, this obstacle can be overcome if we take into account spin-orbit scattering inside the superconductor. Spin-orbit scattering induces a fluctuational (zero on average) spin-triplet component of the superconducting condensate, which is enough to establish Andreev transport into a strong ferromagnet. This remarkably simple mechanism is quite versatile and can explain long-range triplet proximity effect in a number of experimental setups. One particular application of the suggested effect is to measure the spin-orbit scattering time $\tau_{\text{SO}}$ in disordered superconducting materials. The value of Andreev conductance strongly depends on the parameter $\Delta \tau_\text{SO}$ and can be noticeable even in very disordered but relatively light metals like granular aluminum.
\end{abstract}

\date{\today}

\maketitle

\section{Introduction}

Andreev reflection \cite{Andreev64} is a fundamental process underlying electron transport through an interface between a superconductor and a normal metal. An electron excitation incident from the normal side on the surface of a superconductor is reflected as a hole while an extra Cooper pair is added to the superconductor's condensate \cite{PannetierCourtois}. A distinguishing feature of Andreev reflection is that the reflected hole has opposite velocity and spin as compared to the incident electron. Alternatively, the Andreev reflection process can be thought of as a Cooper pair entering from the superconductor into the normal metal and propagating as a coherent state of two electrons bearing the same velocity and opposite spins.

Andreev reflection is crucial for numerous manifestations of the proximity effect \cite{KlapwijkReview, PannetierCourtois} when certain superconducting correlations are observed inside the normal part of a normal metal--superconductor junctions. Most importantly, it explains propagation of the supercurrent in SNS junctions and hence provides microscopic justification of the Josephson effect in such structures.

This simple semiclassical picture implies that Andreev reflection is strongly suppressed in a junction between a superconductor and a metallic ferromagnet \cite{Buzdin05}. In the ferromagnet, electrons with opposite spin projections found themselves in a different microscopic environment and rapidly loose coherence. An important exception from this rule was suggested in Ref.\ \onlinecite{Bergeret01a} (see Ref.\ \onlinecite{Bergeret05} for a review). The main idea is that a coherent pair of electrons in the triplet rather than singlet spin state can much easier propagate inside a ferromagnet when the total spin is aligned with the magnetization direction. It was suggested that a nonuniform ferromagnet (e.g.\ due to the presence of several magnetic domains) can convert an electron pair in the singlet spin state into a triplet provided magnetizations of the domains are not collinear. This mechanism leads to the long-range proximity effect and, in particular, enhanced Josephson current in the SFS junctions \cite{Bergeret01b}.

The suggested mechanism, however, becomes ineffective in extremely strong ferromagnets, often referred to as half-metals, where electrons with only one spin projection can propagate. A singlet Cooper pair simply cannot enter such a material. Nevertheless, there are experiments \cite{HalfMetal1, HalfMetal2} observing a significant triplet supercurrent in half-metallic samples of CrO$_2$ brought in proximity with singlet superconductors. Extremely strong spin polarization of this material was also experimentally confirmed in a related work \cite{HalfMetal3}. As we have already mentioned, appearance of the triplet component in half-metals cannot be explained by the long-range proximity effect of Ref.\ \onlinecite{Bergeret01a} since (i) there is no room for Andreev reflection when only one spin subband is present at the Fermi level and (ii) there is no evidence for noncollinear domains in the experiments.

An alternative mechanism to generate spin-triplet component in a junction between a superconductor and a normal metal or a ferromagnet was suggested in Ref.\ \cite{Nazarov02}. This mechanism involves a special type of boundary between the two materials, such that the electron spin rotates going through the interface. In essence, this is similar to the effect of noncollinear magnetic domains discussed above but in a setting where these magnetic domains are effectively hidden inside the junction interface. Such spin-active boundaries produce a number of non-trivial phenomena \cite{Bergeret12, Eschrig15} and, in particular, can explain Andreev transport between an ordinary singlet superconductor and a fully polarized ferromagnet \cite{Eschrig17}. However, this explanation comes at the cost of assuming that the boundary itself is polarized in a direction not collinear with the magnetization of the ferromagnetic lead.

In this paper we propose an alternative explanation for the superconducting proximity effect in a half-metallic ferromagnet due to possible spin-orbit scattering in the superconducting part of the junction. Spin-orbit scattering naturally occurs in materials composed of heavy elements. Unlike scattering on magnetic impurities, spin-orbit scattering does preserve time-reversal symmetry and hence does not influence the critical temperature of a superconductor \cite{Anderson-SO, AbrikosovGorkov62}. At the same time, it violates spin symmetry and mixes singlet and triplet components of the superconducting condensate. Even in conventional superconductors where Cooper attraction is present only in the singlet channel, Cooper pairs represent a random mixture of singlet and triplet spin states due to spin-orbit scattering. When such a superconductor is brought in contact with the half-metal, triplet Cooper pairs with the suitable orientation of their total spin can easily penetrate the boundary and establish the proximity effect.

To demonstrate this mechanism, we will consider an idealized model of an SIF junction where the superconductor is connected to a half-metallic ferromagnet via a tunnel barrier. We assume the superconductor contains a certain concentration of spin-orbit impurities and has a finite spin-flip scattering time $\tau_\text{SO}$. We will calculate Andreev conductance for arbitrary values of $\Delta\tau_\text{SO}$ and for any conductance of the barrier. In particular, we will demonstrate that in the limit of strong spin-orbit scattering $\Delta\tau_\text{SO} \ll 1$, Andreev conductance has the same order of magnitude as in an equivalent SIN junction, where the ferromagnet is replaced with an ideal normal metallic lead.

Enhancement of Andreev conductance between a half-metallic ferromagnet and a dirty superconductor due to spin-orbit scattering can be used in order to measure spin-orbit scattering time $\tau_\text{SO}$. Such a measurement is expected to be most accurate when the key parameter $\Delta\tau_\text{SO}$ is neither very large nor very small.  Detailed predictions for Andreev conductance as a function
of $\Delta\tau_\text{SO}$ are given in Eqs.\ (\ref{smallGT}), (\ref{largeGT}), and (\ref{general}) and illustrated in Figs.\ \ref{fig:General} and \ref{fig:GDeltatau}  below.

\section{General formalism}

\begin{figure}
  \includegraphics[width=0.9\columnwidth]{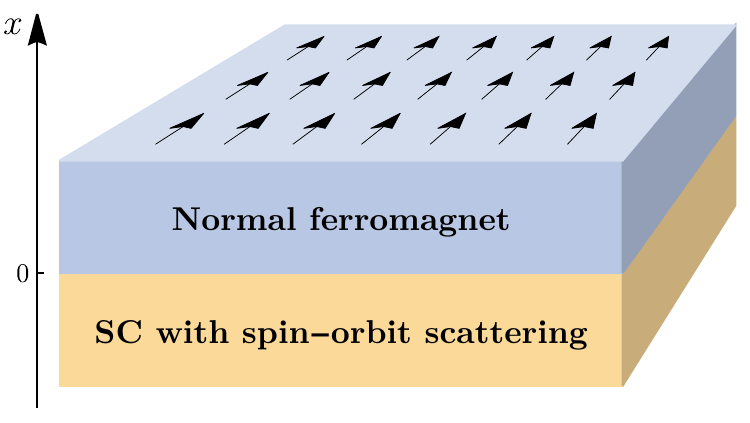}
  \caption{Schematic picture of the SF junction. Arrows on the ferromagnetic layer show its polarization. We assume that the junction is uniform in the latteral dimensions and all physical quantities depend only on the $x$ coordinate.}
  \label{Fig:sample}
\end{figure}

We consider a superconductor-ferromagnet junction shown in Fig.\ \ref{Fig:sample}. Ferromagnetic part occupies the domain $x > 0$ and is assumed clean and ideal (full spin polarization). The superconducting part of the junction at $x < 0$ is in the dirty limit $\Delta\tau \ll 1$ ($\tau$ is the elastic mean-free time) and will be described by the semiclassical Usadel equation \cite{Usadel70}. The superconductor also contains a finite concentration of spin-orbit impurities that induce the spin-flip scattering time $\tau_\text{SO} \gg \tau$. Contact between the superconductor and the half-metal is through the tunnel barrier with the dimensionless conductance $G_T$. This assumption is not very restrictive since the limit of large $G_T$ correctly models a transparent boundary. We assume the system is homogeneous in lateral dimensions and hence it suffices to solve the 1D problem where all quantities depend only on $x$.

Technically, it is more conveninent to study not the Usadel equation but rather an equivalent semiclassical action written in terms of $Q$-matrix. Usadel equation will be ``an equation of motion'' minimizing the action and matrix elements of $Q$ at this minimum coincide with the components of the semiclassical fermionic Green function. The derivation of the semiclassical action is very similar to the derivation of the non-linear sigma model \cite{EfetovBook} but with only one fermionic replica. It is outlined in Appendix \ref{Action}.

The semiclassical action for the semi-infinite superconducting part of the junction has the form
\begin{equation}
 S
  = \frac{\pi\nu}{8} \int_{-\infty}^0 dx \operatorname{Tr} \left[
      D(\nabla Q)^2 - 4\Delta \tau_x Q - \frac{(Q \mathbf{s})^2}{\tau_\text{SO}}
    \right].
 \label{SQ}
\end{equation}
Here $\nu$ is the normal density of states at the Fermi level per one spin component and $D$ is the normal diffusion constant. The matrix $Q$ of the size $8 \times 8$ has the structure in the Nambu, spin, and particle-hole spaces. We denote Pauli matrices operating in these spaces by $\tau$, $s$, and $\sigma$, respectively. The last term of the action contains the vector $\mathbf{s} = \{s_x,\,s_y,\,s_z\}$ of all spin matrices and hence describes isotropic spin-orbit scattering. The matrix $Q$ obeys a nonlinear constraint $Q^2 = 1$ and can be represented as
\begin{equation}
 Q
  = T^{-1} \sigma_z\tau_z T.
 \label{TLambdaT}
\end{equation}
In addition, the symmetry with respect to charge conjugation is also imposed:
\begin{equation}
 Q
  = \bar Q
  \equiv \sigma_x \tau_x s_y Q^T \sigma_x \tau_x s_y.
 \label{QbarQ}
\end{equation}

At the point $x = 0$ the superconductor is connected to a normal ferromagnetic lead that allows propagation of spin-up electrons only (half-metal). We will model this situation by replacing the ferromagnetic part of the junction with a normal metal (both spin projections are allowed and the full spin symmetry is preserved) but assuming that the interface is transparent to the spin-up electrons only. Such a spin-filtering boundary in the tunneling limit is described by the action \cite{Cottet09}
\begin{equation}
 S_\Gamma
  = -\frac{G_T}{8} \operatorname{Tr}[P Q_N Q(0)].
 \label{SGamma}
\end{equation}
Here $G_T$ is the normal conductance of the SF interface per unit area measured in units $e^2/h$ and $P$ is the projection operator that selects the allowed spin component
\begin{equation}
 P
  = \frac{1 + \tau_z s_z}{2}.
 \label{Projector}
\end{equation}
Matrix $Q(0)$ is the value of $Q$ at the interface from the superconducting side. On the opposite ferromagnetic side the matrix is fixed with the value
\begin{equation}
 Q_N
  = \sigma_z \tau_z \cos\chi + \sigma_x \sin\chi.
 \label{QN}
\end{equation}

Let us stress at this point that the boundary action (\ref{SGamma}) couples only spin-up components on the two sides of the junction. This boundary condition does not involve any explicit or hidden noncollinear spin domains and hence is not a spin-active boundary in the sense of Ref.\ \cite{Eschrig15}.

Matrix $Q_N$ on the normal side of the junction contains the source angle $\chi$. Once the total action of the junction is minimized, the linear conductance $G = (dI/dV)_{V\to 0}$ can be computed as (see Appendix \ref{Action} for the derivation)
\begin{equation}
 G
  = \frac{2e^2}{h}\, \frac{\partial^2 S_\text{min}}{\partial\chi^2} \Bigr|_{\chi = 0}.
 \label{GA}
\end{equation}

Since the setup is uniform in the lateral directions, we will consider configurations of $Q$ that depend only on the $x$ coordinate. It is convenient to introduce a dimensionless variable $t = \sqrt{2\Delta/D}\, x$ and a parameter $G_S = 4\pi\nu \sqrt{2D\Delta}$. The latter is the conductance in the normal state in units $e^2/h$ per unit area of a piece of superconductor whose length is equal to the coherence length $\sqrt{D/2\Delta}$. Note that we explicitly include the spin degeneracy factor here. The total action in these dimensionless units is
\begin{multline}
 S
  = \frac{G_S}{32} \int_{-\infty}^0 dt \operatorname{Tr} \left[
      \dot Q^2 - 2 \tau_x Q - \frac{(Q \mathbf{s})^2}{2\Delta \tau_\text{SO}}
    \right] \\ - \frac{G_T}{8} \operatorname{Tr}[P Q_N Q(0)].
 \label{S1}
\end{multline}
This action contains two dimensionless parameters: the strength of spin-orbit scattering $1/\Delta\tau_\text{SO}$ and the ratio $G_T/G_S$ of junction's tunnel conductance to the normal-state conductance of a superconductor. We will first consider two limiting cases of weak contact $G_T \ll G_S$ and of strong spin-orbit scattering $\Delta\tau_\text{SO} \ll 1$. Both cases allow for a relatively simple solution. The general case will be analyzed afterwards.

\section{Two limiting cases}\label{limits}

\subsection{Limit of weak contact}

When the coupling between superconducting and normal parts of the junction is relatively weak, $G_T \ll G_S$, the matrix $Q$ in the superconductor only slightly deviates from its equilibrium bulk value $\tau_x$. Hence we can approximate $Q$ by the expansion
\begin{gather}
 Q = \tau_x (1 + iW - W^2/2), \label{QW} \\
 W = -\bar W,
 \qquad
 \{ \tau_x,\,W \} = 0.
 \label{Wcond}
\end{gather}
Linear constraints on $W$ follow from the conditions $Q^2 = 1$ and $Q = \bar Q$.

Next, we expand the bulk action up to the second order in $W$ and the boundary action up to the linear term:
\begin{multline}
 S
  = \frac{G_S}{32} \int_{-\infty}^0 dt \operatorname{Tr} \left[
      \dot W^2 + W^2 + \frac{3 W^2 - (W \mathbf{s})^2}{2\Delta \tau_\text{SO}}
    \right] \\
    + \frac{G_T}{16} \operatorname{Tr} \Bigl[ (\sigma_z \tau_y \cos\chi + \sigma_x \tau_y s_z \sin\chi) W(0) \Bigr].
\end{multline}
Our goal is to minimize this expression with respect to $W$. From the structure of the boundary term we conclude that only two components in the matrix $W$ are nonzero:
\begin{equation}
 W
  = \sigma_z \tau_y w_s + \sigma_x \tau_y s_z w_t.
 \label{Wweak}
\end{equation}
In terms of these components, the action takes the form
\begin{multline}
 S
  = \frac{G_S}{4} \int_{-\infty}^0 dt \left[
      \dot w_s^2 + \dot w_t^2
      + w_s^2 + \left( 1  +  \frac{2}{\Delta\tau_\text{SO}} \right) w_t^2
    \right] \\
    -\frac{G_T}{2} \bigl[ w_s(0) \cos\chi + w_t(0) \sin\chi \bigr].
 \label{Swswt}
\end{multline}

We observe that $w_s$ and $w_t$ describe soft modes in the singlet and triplet spin sector, respectively. This can be qualitatively explained in the following way. Ferromagnetic part of the junction admits only excitations with a fixed spin projection. A spin state of a fully polarized electron-hole pair $\lvert \uparrow \uparrow \rangle$ can be equivalently viewed as a linear combination of singlet and triplet states $\lvert \uparrow \uparrow \rangle \pm \lvert \downarrow \downarrow \rangle$. (The spin state of an electron-hole pair can be converted into the state of two electrons by applying time reversal to the second spin. Then singlet and triplet acquire a more familiar form $\lvert \uparrow \downarrow \rangle \mp \lvert \downarrow \uparrow \rangle$.) These are exactly the states corresponding to $w_s$ and $w_t$ components in the action (\ref{Swswt}). While the singlet mode is insensitive to spin-orbit scattering, the triplet mode acquires an additional mass $\sim 1/\tau_\text{SO}$.

Minimization of the action (\ref{Swswt}) is straightforward and leads to the result
\begin{align}
 w_s(t)
  &= -\frac{G_T\; e^t}{G_S} \cos\chi,  \\
 w_t(t)
  &= -\frac{G_T\; e^{\sqrt{1 + 2/\Delta\tau_\text{SO}}\, t}}
           {G_S \sqrt{1 + 2/\Delta\tau_\text{SO}}} \sin\chi.
\end{align}
Note that the triplet component $w_t$ vanishes when the source angle $\chi$ is zero. This signifies the absence of the average triplet component of the superconducting Green function in equilibrium.

\begin{figure}
  \includegraphics[width=0.9\columnwidth]{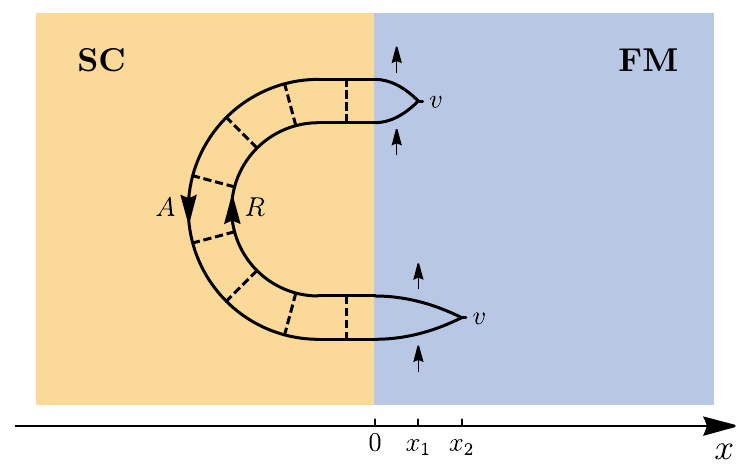}
  \caption{A diagram for Andreev conductance in terms of matrix Green functions for the Bogoliubov -- de Gennes Hamiltonian. The result (\protect\ref{smallGT}) in the limit $G_T \ll G_S$ is provided by just this single diagram. It also illustrates the general Kubo formula (\protect\ref{linearKubo}) and explains the ordering of current operators $0 < x_1 < x_2$ in real space.}
  \label{fig:diffuson}
\end{figure}

The minimized action (\ref{Swswt}) takes the form
\begin{equation}
 S_\text{min}
  = -\frac{G_T^2}{4G_S} \left[ \cos^2\chi + \frac{\sin^2\chi}{\sqrt{1 + 2/\Delta\tau_\text{SO}}} \right]
\end{equation}
while the Andreev conductance is determined by Eq.\ (\ref{GA}) and equals
\begin{equation}
 G
  = \frac{G_T^2}{G_S} \left[ 1 - \frac{1}{\sqrt{1 + 2/\Delta\tau_\text{SO}}} \right].
 \label{smallGT}
\end{equation}
This result can be illustrated by a single diagram shown in Fig.\ \ref{fig:diffuson} with the diffuson ladder in either singlet or triplet spin sector.

In the limit of weak spin-orbit scattering $\tau_\text{SO} \to \infty$, the Andreev conductance vanishes $\propto 1/\tau_\text{SO}$ as expected due to the lack of Andreev reflection in a fully spin-polarized ferromagnet. In the opposite limit $\Delta \tau_\text{SO} \ll 1$, triplet mode is strongly suppressed and the result (\ref{smallGT}) is similar to the usual Andreev conductance of an SIN junction in the tunneling limit \cite{BeenakkerReview}.

\subsection{Strong spin-orbit scattering}

Consider now the limit of very strong spin-orbit scattering $\Delta\tau_\text{SO} \ll 1$ and arbitrary $G_T/G_S$ ratio. The matrix $Q$ becomes trivial in the spin space and we can reduce the problem to the $4 \times 4$ matrices with the action
\begin{equation}
 F
  = \frac{G_S}{16} \int_{-\infty}^0 dt \operatorname{Tr} \left[
      \dot Q^2 - 4 \tau_x Q
    \right] - \frac{G_T}{4} \operatorname{Tr} \bigl[ P Q_N Q(0) \bigr].
\end{equation}
The linear constraint (\ref{QbarQ}) becomes
\begin{equation}
 Q
  = \sigma_x \tau_x Q^T \tau_x \sigma_x.
\end{equation}
This condition defines the standard manifold of the symplectic class sigma model $Q \in \mathrm{O}(4) / \mathrm{O}(2) \times \mathrm{O}(2)$. This manifold has dimension four and is equivalent to the product of two spheres. It can be parametrized explicitly by four angles as
\begin{multline}
 Q
  = [\sigma_z \tau_x \cos\theta_1 + \tau_z \sin\theta_1 \cos\phi_1 + \sigma_z \tau_y \sin\theta_1 \sin\phi_1] \\
    \times [\sigma_z \cos\theta_2 + \sigma_x \tau_z \sin\theta_2 \cos\phi_2 + \sigma_y \tau_z \sin\theta_2 \sin\phi_2].
\end{multline}
In terms of these angles, the action (\ref{S1}) takes the form
\begin{multline}
 S
  = \frac{G_S}{4} \int_{-\infty}^0 dt \Bigl[
      \dot\theta_1^2 + \dot\theta_2^2 \\
      + \sin^2\theta_1 \dot\phi_1^2 + \sin^2\theta_2 \dot\phi_2^2 - 2 \cos\theta_1 \cos\theta_2
    \Bigr] \\
    -\frac{G_T}{2} \sin\theta_1 \cos\phi_1 (\cos\theta_2 \cos\chi + \sin\theta_2 \sin\chi \cos\phi_2) \Bigr|_{t = 0}.
\end{multline}
The bulk part of the action is minimized by any constant values of $\phi_{1,2}$; the boundary term then requires $\phi_{1,2} = 0$. For the remaining two angles, we introduce new variables $\theta_\pm = \theta_1 \pm \theta_2$ and observe that the action decouples in these variables:
\begin{gather}
 S
  = S_U[\theta_+, \chi] + S_U[\theta_-, -\chi], \\
 S_U[\theta, \chi]
  = \frac{G_S}{8} \int_{-\infty}^0 dt \left[ \dot\theta^2 - 2 \cos\theta \right]
    -\frac{G_T}{4} \sin[\theta(0) - \chi].
 \label{SU}
\end{gather}

Minimization of the bulk part of $S_U$ yields the standard Usadel equation for the single angle $\theta$:
\begin{equation}
 \ddot \theta
  = \sin\theta.
 \label{thetaUsadel}
\end{equation}
This equation has an integral of motion and can be reduced to the first order equation
\begin{equation}
 \dot\theta^2 + 2\cos\theta
  = 2,
 \qquad
 \dot\theta = 2 \sin(\theta/2),
 \label{thetaIntegral}
\end{equation}
owing to the fact that $\theta$ should decay in the limit $t \to -\infty$. Using these identities, we can represent the integrand in Eq.\ (\ref{SU}) as a total derivative and express $S_U$ as a function of $\theta$ at the boundary:
\begin{equation}
 S_U
  = -G_S \cos\frac{\theta(0)}{2} - \frac{G_T}{4} \sin \bigl[ \theta(0) - \chi \bigr].
 \label{SUt0}
\end{equation}

In the absence of the source field, $\chi = 0$, minimum of $S_U$ is attained at:
\begin{equation}
 \sin\frac{\theta_0}{2} = \frac{\sqrt{G_S^2 + 2 G_T^2} - G_S}{2 G_T}.
 \label{theta0}
\end{equation}
A small nonzero value of $\chi$ can be then taken into account perturbatively. We expand up to the second order in $\chi$ and then apply Eq.\ (\ref{GA}). This yields the following result:
\begin{equation}
 G
  = \frac{G_S \left( \sqrt{G_S^2 + 2 G_T^2} - G_S \right)^{3/2} \sqrt{\sqrt{G_S^2 + 2 G_T^2} + 3G_S}}{2 G_T \sqrt{G_S^2 + 2 G_T^2}}.
  \label{strongSO}
\end{equation}

As was already mentioned previously, strong spin-orbit scattering completely suppresses triplet electron modes in the superconductor. Hence the above result for Andreev conductance is solely due to semiclassical dynamics of singlet electron-hole pairs. One quite remarkable consequence of this observation is that an SIN junction with a tunnel barrier but without ferromagnetism exhibits exactly the same dependence (\ref{strongSO}) of Andreev conductance on $G_S$ and $G_T$. The only minor distinction is that $G_T$ in that case should include contributions from both spin-up and spin-down conducting channels through the barrier and hence is twice larger than in an equivalent SF junction. One other crucial assumption that allows to equally apply Eq.\ (\ref{strongSO}) to both SIN and SF junctions is that the normal part should be relatively clean compared to the superconducting side of the junction.

In the two limiting cases of weak and strong barrier, we can reduce Eq.\ (\ref{strongSO}) to
\begin{equation}
 G
  = \begin{dcases}
      \frac{G_T^2}{G_S}, & G_T \ll G_S, \\
      \frac{G_S}{\sqrt{2}}, &  G_T \gg G_S.
    \end{dcases}
 \label{strongSOlimits}
\end{equation}
The first of these cases matches with the result (\ref{smallGT}) when $\Delta\tau_\text{SO} \ll 1$. The second limit corresponds to usual Andreev conductance in a dirty SN junction with a relatively transparent interface as was explained above.

\section{Andreev conductance at arbitrary \texorpdfstring{$\bm{\Delta\tau_\text{SO}}$}{\textDelta\texttau SO} and interface quality}

\subsection{General result}

When both dimensionless parameters of the problem $\Delta\tau_\text{SO}$ and $G_T/G_S$ take arbitrary values we can apply the following strategy to find Andreev conductance. First, minimize the action in the absence of the source field, $\chi = 0$. It turns out that spin-orbit scattering is irrelevant in this case and the action is minimized by the simple one-parameter trajectory in the singlet sector only:
\begin{equation}
 Q
  = \tau_x e^{-i \sigma_z \tau_y \theta(t)}.
 \label{Qtheta}
\end{equation}
The action for this single angle $\theta$ has the standard Usadel form:
\begin{equation}
 S_0
  = \frac{G_S}{4} \int_{-\infty}^0 dx \bigl[ \dot\theta^2 - 2 \cos\theta \bigr]
    - \frac{G_T}{2} \sin\theta(0).
 \label{Stheta}
\end{equation}
This action is twice larger than Eq.\ (\ref{SU}) in the limit $\chi = 0$.

The function $\theta(t)$ can be found by solving the Usadel equation (\ref{thetaUsadel}). Using the integral of motion (\ref{thetaIntegral}), we obtain
\begin{equation}\label{theta}
 \theta(t)
  = 4 \arctan e^{t - t_0},
 \qquad
 t_0
  = -\ln\tan\frac{\theta(0)}{4}.
\end{equation}
Substituting this result back into Eq.\ (\ref{Stheta}), we express the action as a function of $\theta(0)$,
\begin{equation}\label{zeroaction}
 S_0
  = -2 G_S \cos\frac{\theta(0)}{2} - \frac{G_T}{2} \sin\theta(0).
\end{equation}
Similarly to Eq.\ (\ref{Stheta}), this expression is twice larger than Eq.\ (\ref{SUt0}) provided $\chi = 0$. Hence the minimum of the action is attained when $\theta(0)$ is given by Eq.\ (\ref{theta0}).

Next step is to expand the action in small deviations in the vicinity of the solution (\ref{Qtheta}) and at the same time to take into account a small but nonzero value of the source angle $\chi$. This is achieved by representing $Q$ as
\begin{equation}
 Q
  = e^{i \sigma_z \tau_y \theta(t)/2} \tau_x (1 + iW - W^2/2) e^{-i \sigma_z \tau_y \theta(t)/2},
 \label{QthetaW}
\end{equation}
where $W$ obeys conditions (\ref{Wcond}). We expand the bulk action up to the second order in $W$ and the boundary action up to the second order in either $W(0)$ or $\chi$. This procedure is described in detail in Appendix \ref{General}. Minimizing the quadratic action in $W$ at a given $\chi$ and applying Eq.\ (\ref{GA}) yields the following result:
\begin{multline}\label{general}
 G
  = G_T \sin 2\theta_0 \biggl[
      4 + 3\Delta\tau_\text{SO} - 2\cos\theta_0 \\
      + \frac{4 - 6\Delta\tau_\text{SO}}
        {3 + 2\sqrt{1 + 2/\Delta\tau_\text{SO}} \cos(\theta_0/2) + \cos\theta_0}
    \biggr]^{-1}.
\end{multline}

\begin{figure}
  \includegraphics[width=0.9\columnwidth]{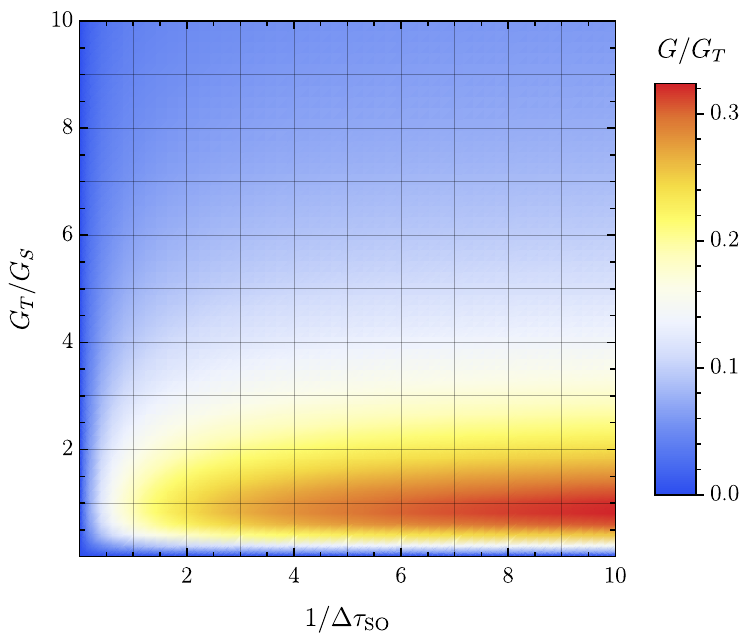}\\
  \includegraphics[width=0.9\columnwidth]{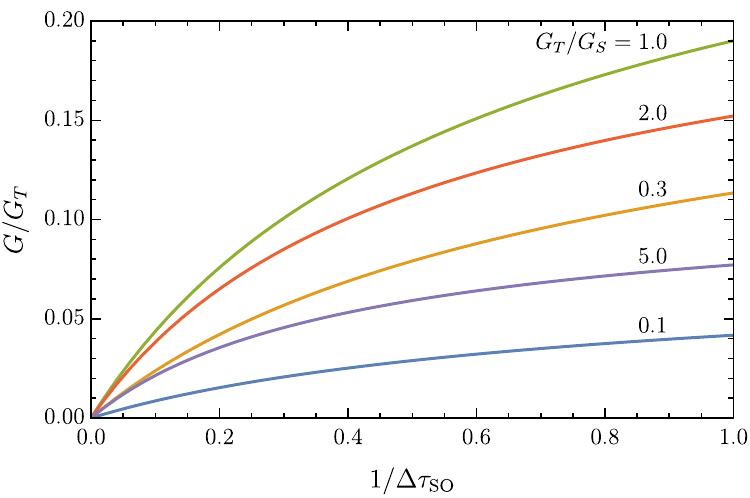}\\
  \includegraphics[width=0.9\columnwidth]{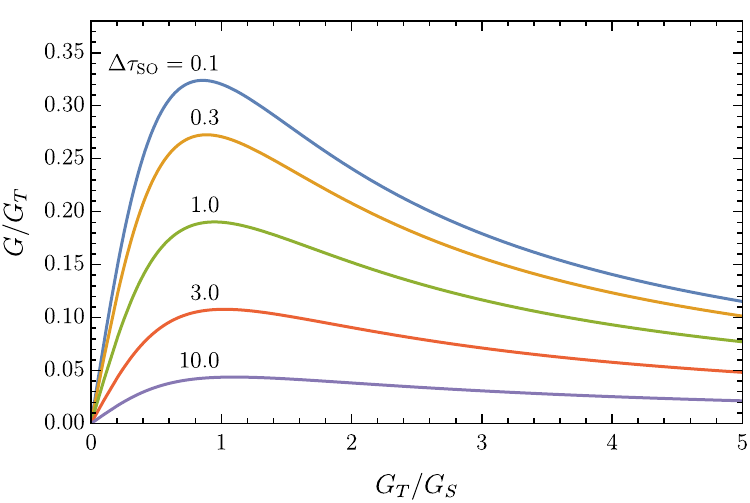}
  \caption{Andreev conductance normalized to $G_T$ as a function of both spin-orbit scattering rate $1/\Delta\tau_\text{SO}$ and conductance ratio $G_T/G_S$ according to Eq.\ (\protect\ref{general}) (top). The same dependence for several fixed values of $G_T/G_S$ (middle) and $\Delta\tau_\text{SO}$ (bottom).}
  \label{fig:General}
\end{figure}

We have thus established a general expression for the linear Andreev conductance valid for arbitrary values of parameters $\Delta\tau_\text{SO}$ and $G_T/G_S$. The only assumptions are relatively large (compared to both $G_T$ and $G_S$) conductance of the ferromagnetic part of the junction and low temperature $T \ll \Delta$. Dependence of Andreev conductance on both parameters is illustrated in Fig.\ \ref{fig:General}.

Let us stress once more that spin-orbit scattering in the superconductor is crucial for Andreev transport in the SF junction. Andreev conductance is an increasing function of spin-orbit scattering rate $1/\Delta\tau_\text{SO}$ and vanishes when this parameter is zero, see Fig.\ \ref{fig:General} (middle panel). Andreev conductance is also suppressed both when the junction coupling is weak, $G_T \ll G_S$, and in the limit $G_S \ll G_T$ when superconductivity is weak. Highest values of Andreev conductance are achieved at $G_T \approx G_S$ as can be seen in Fig.\ \ref{fig:General} (bottom panel).

\subsection{Analysis of limiting cases}

Naturally, the general expression (\ref{general}) for Andreev conductance is consistent with the limiting cases considered previously in Sec.\ \ref{limits}. The limit of weak junction $G_T \ll G_S$ implies small values of the angle $\theta_0 \approx G_T/G_S \ll 1$ according to Eq.\ (\ref{theta0}). In this limit, Eq.\ (\ref{general}) directly reduces to Eq.\ (\ref{smallGT}).

\begin{figure}
  \includegraphics[width=0.9\columnwidth]{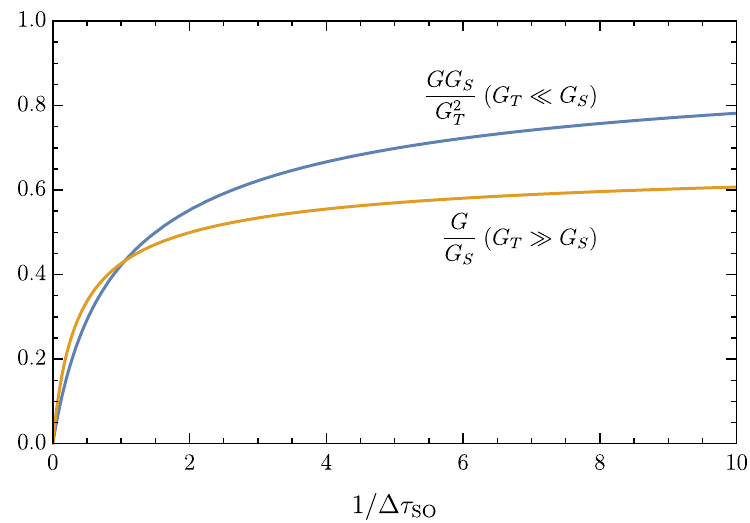}
  \caption{Dependence of Andreev conductance on the spin-orbit scattering rate $1/\Delta\tau_\text{SO}$ in the limits of weak and strong coupling in the SF junction. Upper curve for the case $G_T \ll G_S$ corresponds to Eq.\ (\protect\ref{smallGT}). Lower curve illustrates the dependence Eq.\ (\protect\ref{largeGT}) for $G_T \gg G_S$. The plots in Fig.\ \protect\ref{fig:General} (middle panel) interpolate between these two limits.}
  \label{fig:GDeltatau}
\end{figure}

In the opposite case $G_T \gg G_S$, the value of $\theta_0$ is close to $\pi/2$ and Andreev conductance acquires the form
\begin{equation}
 G
  = \frac{G_S}{\sqrt{2}} \biggl[
      1 + \frac{3}{4}\Delta\tau_\text{SO}
      + \frac{1 - (3/2)\Delta\tau_\text{SO}}
        {3 + \sqrt{2 + 4/\Delta\tau_\text{SO}}}
    \biggr]^{-1}.
 \label{largeGT}
\end{equation}
Remarkably, this result is independent of the value of $G_T$. Both limiting functions are shown in Fig.\ \ref{fig:GDeltatau}.

\begin{figure}
  \includegraphics[width=0.9\columnwidth]{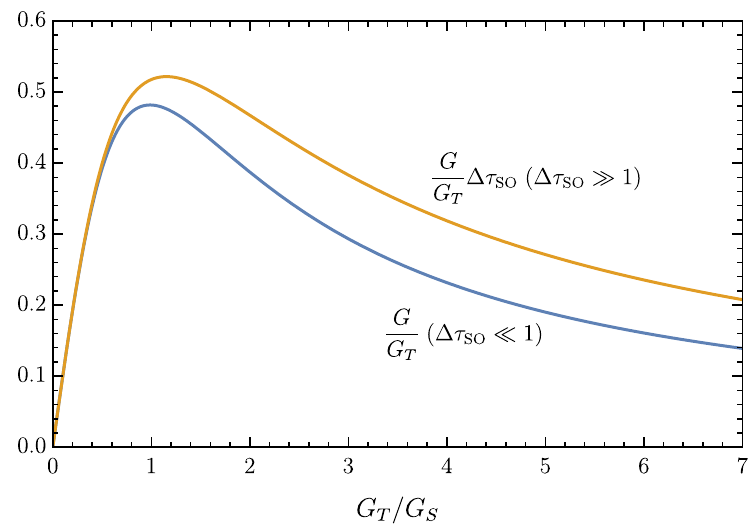}
  \caption{Andreev conductance as a function of $G_T/G_S$ in the limits of weak [upper curve, Eq.\ (\protect\ref{weakSO})] and strong [lower curve, Eq.\ (\protect\ref{strongSO})] spin-orbit scattering. The plots in Fig.\ \protect\ref{fig:General} (lower panel) interpolate between these two limits.}
  \label{fig:GGTGS}
\end{figure}

Another pair of asymptotic results is for the limits of weak and strong spin-orbit scattering. In the latter case, $\Delta\tau_\text{SO} \ll 1$, we can expand Eq.\ (\ref{general}) and obtain
\begin{equation}
 G
  = \frac{G_T \sin 2\theta_0}{4 - 2\cos\theta_0}
    -\frac{\sqrt{2\Delta\tau_\text{SO}}\, G_T \sin 2\theta_0}{(4 - 2\cos\theta_0)^2\cos(\theta_0/2)} + \ldots
\end{equation}
The first term of this expansion reproduces Eq.\ (\ref{strongSO}) upon substitution of Eq.\ (\ref{theta0}). The second term represents a small correction $\sim \sqrt{\Delta\tau_\text{SO}}$ to this result and takes a surprisingly simple form
\begin{equation}
 \delta G
  = -\frac{\sqrt{\Delta\tau_\text{SO}} G_T^2 G_S}{\sqrt{2} (G_S^2 + 2G_T^2)}.
\end{equation}
When spin-orbit scattering is weak, $\Delta\tau_\text{SO} \gg 1$, Eq.\ (\ref{general}) reduces to
\begin{multline}
 G
  = \frac{4 G_T \cos\theta_0 [1 - \cos^3(\theta_0/2)]}{3 \Delta \tau_\text{SO} \sin(\theta_0/2)}
  = \frac{2\sqrt{2} G_S}{3 G_T^3 \Delta \tau_\text{SO}} \Biggl[
      2\sqrt{2} G_T^3 \\
      - \left( G_S \sqrt{G_S^2 + 2G_T^2} - G_S^2 + G_T^2 \right)^{3/2}
    \Biggr].
 \label{weakSO}
\end{multline}
This dependence, together with Eq.\ (\ref{strongSO}), is shown in Fig.\ \ref{fig:GGTGS}.

Finally, we quote the results for simultaneous limits when both the spin-orbit scattering and the junction coupling are weak or strong:
\begin{equation}
 G
  = \begin{dcases}
      \frac{G_T^2}{G_S}, & G_T \ll G_S, \quad \Delta\tau_\text{SO} \ll 1, \\
      \frac{G_S}{\sqrt{2}}, & G_T \gg G_S, \quad \Delta\tau_\text{SO} \ll 1, \\
      \frac{G_T^2}{G_S \Delta\tau_\text{SO}}, & G_T \ll G_S, \quad \Delta\tau_\text{SO} \gg 1, \\
      \frac{8 - 2\sqrt{2}}{3} \frac{G_S}{\Delta\tau_\text{SO}}, & G_T \gg G_S, \quad \Delta\tau_\text{SO} \gg 1.
    \end{dcases}
\end{equation}
The first two cases here naturally coincide with Eq. (\ref{strongSOlimits}).

\section{Discussion and conclusions}

In the present paper we have demonstrated that Andreev conductance between a strong half-metal ferromagnet (like CrO$_2$ for example) and a dirty superconductor can be of significant magnitude due to spin-orbit scattering in the superconductor. Our key qualitative observation is as follows:  the triplet component of the Cooper pair wave function (necessary for this conduction mechanism to operate) is not required to have a nonzero spatial average. In fact, it is sufficient to have a finite \textit{square average} of the triplet component. This is exactly what spin-orbit scattering in a singlet superconductor generates.

We have found Andreev conductance between a dirty superconductor with the order parameter $\Delta$ and spin-orbit scattering rate $1/\tau_\text{SO}$ and an idealized ferromagnet connected by a generic tunnel junction with the conductance $G_T$. The result is a function of two dimensionless parameters $\Delta\tau_\text{SO}$ and $G_T/G_S$ where $G_S = 4\pi\nu \sqrt{2D\Delta}$ is the conductance in the normal state of a piece of superconductor whose length is equal to the coherence length $\sqrt{D/2\Delta}$. Our general result for Andreev conductance is given in Eqs.\ (\ref{general}) and (\ref{theta0}) and illustrated in Fig.\ \ref{fig:General}. Calculation of the Andreev conductance in two limits of weak tunneling coupling $G_T \ll G_S$ or strong spin-orbit scattering $\Delta\tau_\text{SO} \ll 1$ is technically easier compared to the general case. This calculation is presented first with the results in Eqs.\ (\ref{smallGT}) and (\ref{strongSO}), respectively. Opposite limits of strong tunneling coupling $G_T \gg G_S$ or weak spin-orbit scattering $\Delta\tau_\text{SO} \gg 1$ are derived later from the general expression (\ref{general}) and given by Eq.\ (\ref{largeGT}) and (\ref{weakSO}). All four limiting forms of the Andreev conductance are shown in Figs.\ \ref{fig:GDeltatau} and \ref{fig:GGTGS}.

Quite remarkably, Andreev conductance of the SIF junction with strong spin-orbit scattering has exactly the same functional dependence (\ref{strongSO}) on $G_T$ and $G_S$ as the conductance of an equivalent SIN junction without both ferromagnetic order and spin-orbit scattering.

In general, spin-orbit scattering rate can be estimated as $1/\tau_\text{SO} \approx (Z/137)^4 / \tau$ where $\tau$ is the electron mean-free time and $Z$ is the typical atomic number of the material. The effect proposed in the present paper is easier to observe experimentally in relatively dirty superconductors in the presence of heavy elements. For example, for amorphous indium oxide InO$_x$ with $Z_\text{In} = 49$ and $\tau \sim 3\cdot 10^{-16}$s one has $\tau_\text{SO} \sim 2\cdot 10^{-14}$s which results in a very small product $\Delta\tau_\text{SO} \sim 10^{-2}$. This estimate may explain high values of the upper critical field $H_{c2}$ observed in Ref.\ \cite{Sacepe2015} which strongly exceed the usual Chandrasekhar-Clogston paramagnetic limit $H_P = \Delta/\sqrt{2}\mu_B$. Indeed, short spin-orbit scattering time leads \cite{Anderson-SO} to enhancement of paramagnetic critical field according to the relation $H_P^\text{SO} \sim H_P/\sqrt{\Delta\tau_\text{SO}}$.

We predict that Andreev conductance between amorphous InO$_x$ and half-metallic ferromagnet CrO$_2$ should be of the same order of magnitude as between InO$_x$ and some simple normal metal without spin polarization.  Dependence of Andreev conductance on the basic parameters $G_T$ and $G_S$ in both cases is given by Eq.\ (\ref{strongSO}).

The above example shows that considerable spin-orbit scattering rate with $\Delta\tau_\text{SO} \sim 1$ can be found in other superconducting materials, even composed of lighter elements, if the potential disorder is strong and hence the elastic scattering time is short enough. For example, for highly resistive granular aluminum with $Z_\text{Al} = 13$ it is quite realistic to obtain $\hbar/\tau_\text{SO} \sim 10^{-4}$eV comparable to the superconducting gap in Al. Then the actual magnitude of $\tau_\text{SO}$ can be found by comparing measured Andreev conductivity from a half-metallic ferromagnet with our theory. Namely, for a weak interface coupling $G_T \ll G_S$ we predict the dependence of Andreev conductance on $\tau_\text{SO}$ given by Eq.\ (\ref{smallGT}). In a general case of arbitrary $G_T/G_S$ ratio, one should use Eq.\ (\ref{general}). On the other hand, relatively clean superconductors with $\Delta\tau \gtrsim 1$ cannot provide necessary spin-orbit scattering since $\tau_\text{SO} \gg \tau$ always. In this case the triplet component is very small and the magnitude of sub-gap Andreev conductance will rather indicate the amount of spin polarization in the ferromagnet \cite{HalfMetal1}.

Our results were derived in the limit of zero temperature $T$ and zero voltage $V$ (linear regime). Practically, it means that we assume $T, eV \ll \Delta$.  Although it is possible to generalize our calculations for nonzero $T$ and $V$, we defer this to a separate publication. The same concerns the account of non-complete polarization of the ferromagnet. Indeed, if the minority spin sub-band is not totally suppressed, so that its contribution to the density of states at the Fermi energy is some small fraction $\alpha_\text{FM} = \nu_\downarrow/\nu_\uparrow \ll 1$, then a weak Andreev conductance can exist even without the triplet component and in the absence of spin-orbit scattering. This restricts the possibility to measure by the proposed method very long spin-orbit times $\tau_\text{SO} \gtrsim 1/\alpha_\text{FM} \Delta$.

Josephson effect in an SFS junction with a strong ferromagnet \cite{HalfMetal2} represents another example of coherent transport which is controlled by the triplet component of the pair amplitude. While we postpone an actual calculation of the Josephson critical current in such a device to a separate publication, our major point about the role of spin-orbit scattering for Andreev conductance is relevant in the case of Josephson current as well. Usual explanation \cite{Bergeret01a, Bergeret01b, Bergeret05} for long-range Josephson current in terms of the triplet component of pairing as a result of non-collinear magnetization is hardly applicable in the case of fully polarized materials like CrO$_2$. On the other hand, superconducting alloy NbTiN studied in Ref.\ \cite{HalfMetal2} is sufficiently disordered to provide spin-orbit scattering rate $1/\tau_\text{SO}$ comparable to $\Delta$ and to realize the mechanism of coherent pair transfer proposed in our paper.

\section{Acknowlegments}

We are grateful to Yakov Fominov for fruitful discussions at the initial stage of this project and to Teunis Klapwijk for valuable comments on the manuscript.

\appendix

\section{\label{Action} Derivation of the effective action}

In this Appendix we outline the derivation of the action (\ref{S1}) together with the source terms that define the matrix $Q_N$ in Eq.\ (\ref{QN}). We assume the contact has the form of a one-dimensional wire characterized by a set of conducting channels. Kinetic energy of the electrons in the metallic wire is given by the operator $\xi = v p$, where $v$ is the velocity operator acting in the space of conducting channels and $p = -i(\partial/\partial x)$. Together with random scalar potential $U$ and random spin-orbit scattering amplitude $U_\text{SO}$, single-particle Hamiltonian takes the following form:
\begin{equation}
 h = v p + U + U_\text{SO},
 \qquad
 h = s_y h^* s_y.
\end{equation}
The second identity here is the time-reversal symmetry of the model.

In the superconducting state, the system is described by the Bogoliubov -- de Gennes Hamiltonian, that obeys an additional mirror symmetry,
\begin{equation} \label{BdG}
 H = \begin{pmatrix}
       h & \Delta \\
       \Delta^* & -h
     \end{pmatrix},
 \qquad
 H = - \tau_y s_y H^* \tau_y s_y.
\end{equation}
We will describe the normal part of the junction by the same Hamiltonian assuming both $\Delta$ and $U_\text{SO}$ are absent while potential disorder $U$ is relatively weak. In other words, we assume the normal metallic lead is equivalent to a clean ballistic waveguide for electron waves. In our model, the tunnel interface between the ideal normal and dirty superconducting sides of the junction is transparent for spin-up electrons only.

\subsection{Kubo formula}

Andreev conductance can be obtained using the approach of Blonder, Tinkham, and Klapwijk \cite{BTK} in terms of probability of Andreev reflection in individual conducting channels. This probability is then expressed via Green functions of the system leading to the Kubo formula \cite{Nazarov94}. In the regime of linear response, Andreev conductance is
\begin{equation}\label{KuboT}
 G
  = -\frac{e^2}{h} \int_0^\infty dE \frac{\partial f_0}{\partial E} \operatorname{Tr} \bigl[v G^R_E(x_1, x_2) v G^A_E(x_2, x_1) \bigr].
\end{equation}
Here $f_0(E)$ is the Fermi distribution function, $x_{1,2} > 0$ are two arbitrary points within the normal side of the junction, and the trace is taken both in the Nambu and channel spaces. Detailed derivation of this formula will be published elsewhere.

Retarded and advanced Green functions are defined in the standard way:
\begin{equation}
 G^{R/A}_E = (E - H \pm i0)^{-1}.
\end{equation}
They are matrices in the Nambu and channel spaces. Retarded and advanced Green functions are related by the symmetry (\ref{BdG}):
\begin{equation}
 G^A_E(x_1, x_2) = -\tau_y s_y \left[ G^R_{-E} \right]^T(x_2, x_1) \tau_y s_y.
 \label{GRGA}
\end{equation}

In the limit of small temperature $T \ll \Delta$, the Kubo formula greatly simplifies. Derivative of the Fermi distribution function can be replaced with a delta-function of energy hence integration in Eq.\ (\ref{KuboT}) is removed.
\begin{equation}
 G
  = \frac{e^2}{2h} \operatorname{Tr} \left[ v G^R(x_1, x_2) v G^A(x_2, x_1) \right]_{E = 0}.
 \label{linearKubo}
\end{equation}
In the rest of the paper, we consider only this limit. The Kubo formula (\ref{linearKubo}) is illustrated in Fig.\ \ref{fig:diffuson}.

The formula for Andreev conductance can be further expressed as a Gaussian path integral over fermionic fields with the action
\begin{equation}
 S
  = -i \int dx\, \phi^\dag (i0 - H) \phi.
 \label{SGR}
\end{equation}
The field $\phi$ is a column vector in the Nambu and channel spaces whose elements are Grasmann-valued functions of $x$. The field $\phi^\dag$ is an independent row vector of a similar structure. The action (\ref{SGR}) is written in such a way that it generates the retarded Green function at zero energy. This is sufficient for the problem at hand due to the identity (\ref{GRGA}). Let us note that in order to solve the problem at a finite temperature we would need a more general model with the action that encodes both retarded and advanced Green functions at the same nonzero energy independently.

The Andreev conductance (\ref{linearKubo}) can be written as the following correlation function of four fields:
\begin{equation}
 G
  =  \frac{e^2}{4h} \left< \Bigl( \phi^T s_y \tau_y v \phi \Bigr)_{x_1} \Bigl( \phi^\dag v \tau_y s_y \phi^* \Bigr)_{x_2} \right>.
 \label{Gphi}
\end{equation}
Here angular brackets imply averaging with the Gaussian weight $e^{-S}$ with the action (\ref{SGR}). Indeed, applying the Wick theorem to the product of four fields in Eq. (\ref{Gphi}), we get two terms, both with a product of two retarded Green functions. Together, they can be combined as
\begin{multline}
 G
  = \frac{e^2}{4h} \operatorname{Tr} \Bigl[ s_y \tau_y v G^R(x_1, x_2) \\
    \times(\tau_y s_y v^T - v \tau_y s_y) (G^R)^T(x_1, x_2) \Bigr].
 \label{GWick}
\end{multline}
Finally, using the identity (\ref{GRGA}) and the time-reversal property of the velocity operator $v^T = - s_y v s_y$, we indeed reproduce the expression (\ref{linearKubo}).

Let us stress that formula (\ref{Gphi}) is valid only for a system with broken spin symmetry (symplectic class). When the spin symmetry is preserved time-reversal operation does not involve $s_y$ matrix and the two terms from the Wick theorem in Eq.\ (\ref{GWick}) cancel each other. In this case, a similar correlation function for the Andreev conductance can be written in terms of commuting rather than Grassmann fields.

\subsection{Derivation of the source term}

We can further transform the action (\ref{SGR}) taking into account superconducting symmetry (\ref{BdG}) and introducing the particle-hole space. We also include the source terms in the action.
\begin{equation}
 S
  = -\frac{i}{2} \begin{pmatrix} \phi^\dag, & \phi^T s_y \tau_y \end{pmatrix}
    \begin{pmatrix} i0 - H & z_+ v \\ z_- v & -i0 - H \end{pmatrix}
    \begin{pmatrix} \phi \\ \tau_y s_y \phi^* \end{pmatrix}.
 \label{Szz}
\end{equation}
When $z_\pm = 0$, this action is indeed identical to Eq.\ (\ref{SGR}). Andreev conductance in the form (\ref{Gphi}) can be generated as a variation of the partition function:
\begin{equation}
 G
  = -\frac{e^2}{h Z} \frac{\partial^2 Z}{\partial z_-(x_1) \partial z_+(x_2)},
 \qquad
 Z
  = \int d\phi^* d\phi\; e^{-S}.
 \label{GZ}
\end{equation}

In order to average Andreev conductance with respect to disorder, we should get rid of $Z$ in the denominator of the above equation. This is usually achieved with the help of the replica trick that extends the system, and hence the fields, to $N$ identical copies. Alternatively, one can apply supersymmetry formalism \cite{EfetovBook}, which augments the fermionic model with its bosonic counterpart. Both approaches lead to the identity $Z = 1$ in the extended theory. We will ignore these complications since our final goal is to derive the Usadel equation rather than a complete field theory. Usadel equation provides a fully symmetric (in replicas or superspace) minimum of the effective action. Hence we can as well proceed with the derivation using the original action (\ref{Szz}) and simply disregard $Z$ in the denominator in Eq.\ (\ref{GZ}).

The derivation of the effective action proceeds in the standard way \cite{EfetovBook}. We introduce notations for doubled fields in the PH space and rewrite the action (\ref{Szz}) as
\begin{gather}
 S
  = -i \bar\psi \Bigl[ i0 \sigma_z \tau_z - \tau_z H + \tau_z \hat v (z_- \sigma_- + z_+ \sigma_+) \Bigr] \psi, \label{Spsi} \\
 \psi
  = \frac{1}{\sqrt{2}} \begin{pmatrix} \phi \\ \tau_y s_y \phi^* \end{pmatrix}, \quad
 \bar\psi
  = \frac{1}{\sqrt{2}} \begin{pmatrix} \phi^\dag \tau_z, & i \phi^T s_y \tau_x \end{pmatrix}. \label{psibarpsi}
\end{gather}
Here $\sigma_z$ and $\sigma_\pm = \sigma_x \pm i \sigma_y$ are Pauli matrices in the PH space.

The partition function (\ref{GZ}) is then averaged with respect to the random Gaussian potential $U$ and the ensuing quartic term is decoupled by a subsequent Hubbard-Stratonovich transformation. This introduces a new matrix field in the problem, which we denote $\tilde Q$, and the action becomes
\begin{equation}
 S
  = \frac{\pi\nu}{8\tau} \operatorname{Tr} \tilde Q^2
    -i \bar\psi \left[ \frac{i\tilde Q}{2\tau} - \tau_z H_0 + \tau_z \hat v (z_- \sigma_- + z_+ \sigma_+) \right] \psi.
 \label{SQpsi}
\end{equation}
Here $H_0$ is the Hamiltonian (\ref{BdG}) without random potential and $\tilde Q$ obeys the symmetry condition (\ref{QbarQ}). The latter follows from the fact that $\bar\psi$ and $\psi$ are linearly related to each other, cf.\ Eq.\ (\ref{psibarpsi}).

In the normal part of the junction, $x > 0$, $H_0$ reduces to the clean metallic Hamiltonian $H_0 = \tau_z v p$. This allows us to exclude the source terms from the action by applying a suitable gauge rotation:
\begin{equation}
 Q
  = M^{-1} \tilde Q M,
 \qquad
 M^{-1}
  = \bar M.
 \label{gauge}
\end{equation}
The matrix $M$ should be chosen such that
\begin{equation}
 \frac{\partial M}{\partial x}
  = i \tau_z (z_- \sigma_- + z_+ \sigma_+) M.
 \label{Mevol}
\end{equation}
The source terms $z_\mp$ are only relevant at two spatial points $x_{1,2}$ according to Eq.\ (\ref{GZ}). We will thus assume $z_\mp(x)$ to be proportional to delta functions $\delta(x - x_{1,2})$. In fact, the partition function $Z$ depends only on the product of amplitudes of these two delta functions hence we can further introduce a single source angle $\chi$ by the relation
\begin{equation}
 z_\mp(x) = \sin(\chi/2) \delta(x - x_{1,2}).
\end{equation}
With this definition, Eq.\ (\ref{GZ}) directly reduces to Eq.\ (\ref{GA}) used in the main text of the paper (up to the factor $Z$ in the denominator, which we discard in the semiclassical limit).

The result is also independent of the exact positions of $x_{1,2}$. For the sake of simplicity, we will assume the ordering $0 < x_1 < x_2$ and take the limit $x_{1,2} \to 0$, as shown in Fig.\ \ref{fig:diffuson}. With these assumptions, Eq.\ (\ref{Mevol}) is readily solved yielding a step-like matrix $M$
\begin{equation}
 M_{x > 0}
  = \begin{pmatrix} 1 & i \tau_z \sin(\chi/2) \\ i \tau_z \sin(\chi/2) & \cos^2(\chi/2) \end{pmatrix} M_{x < 0}.
 \label{MM}
\end{equation}

Once the gauge transformation has removed the source term from Eq.\ (\ref{SQpsi}), derivation of the effective action proceeds in the standard way \cite{EfetovBook}. Fermionic fields $\psi$ are integrated out, and the resulting nonlinear action in terms of $Q$ is restricted to its saddle manifold defined by Eqs.\ (\ref{TLambdaT}) and (\ref{QbarQ}). Expansion of the action in small gradients of $Q$ as well as in small parameters $\Delta\tau$ and in spin-orbit scattering amplitude $U_\text{SO}$ yields the action (\ref{SQ}). It only remains to establish boundary conditions at infinity and at the interface. This is where the source terms show up.

\subsection{Boundary conditions}

Far in the normal part of the junction, $x \to +\infty$, we can completely disregard any proximity effect. In this limit the matrix field attains its limiting value $\tilde Q = \sigma_z \tau_z$. This value is fixed by the infinitely small term in Eq.\ (\ref{Spsi}) to prevent exponential growth of $\psi$. Since we assume the normal metal to be almost clean, its diffusion constant is large and gradients of $Q$ are strongly suppressed. Hence we can simply set $\tilde Q = \sigma_z \tau_z$ everywhere in the normal lead. For the matrix $Q$, this implies
\begin{equation}
 Q_N = Q_{x > 0} = M^{-1}_{x > 0} \sigma_z \tau_z M_{x > 0}.
\end{equation}

Deep in the superconducting part of the junction $x < 0$, we have an additional term in the Hamiltonian with $\Delta$. This term acts in the Nambu space only and we would like to preserve its form under the gauge transformation (\ref{gauge}). This means we should assume $M_{x < 0}$ to act trivially in the Nambu space. We will thus choose a diagonal matrix
\begin{equation}
 M_{x < 0}
  = \begin{pmatrix}
      i\cos(\chi/2) & 0 \\ 0 & 1
    \end{pmatrix}.
 \label{M0}
\end{equation}
With this choice, Eqs.\ (\ref{MM}) and (\ref{M0}) yield the simplest possible form of $Q_N$:
\begin{equation}
 Q_N
  = \begin{pmatrix}
      \tau_z \cos\chi & \sin\chi \\ \sin\chi & -\tau_z \cos\chi
    \end{pmatrix}.
\end{equation}
This is exactly the boundary value (\ref{QN}) used in the main text of the paper.

The last ingredient of the model is the boundary action (\ref{SGamma}) at the interface $x = 0$. This part of the action can be derived in its most general form \cite{EfetovBook, Nazarov01, Cottet09, Eschrig15}
\begin{equation}
 S_\Gamma
  = -\frac{1}{4} \operatorname{Tr} \ln[1 + \mathcal{T} Q(0) \mathcal{T}^\dag Q_N].
\end{equation}
Here $\mathcal{T}$ is the transfer matrix of the boundary \cite{BeenakkerReview}. This matrix is twice larger than the space of channels (it additionally has the left/right structure) and has eigenvalues $e^{\pm \lambda_n}$ related to the transmission probabilities of individual channels $T_n = \cosh^{-2}\lambda_n$. In our problem, $\mathcal{T}$ is trivial in the PH space but does discriminate between up and down spins and hence does not commute with $Q$. Nevertheless, we can choose a specific basis in the channel and spin spaces to bring $\mathcal{T}$ to its diagonal form. Assuming all $\lambda_n$ for spin-down channels are very large (negligible transparency of the interface), we can reduce the boundary action to
\begin{equation}
 S_\Gamma
  = -\frac{1}{2} \operatorname{Tr} \ln[1 + e^{-2\lambda_n} P Q(0) P Q_N].
\end{equation}
Here $P$ is the projection operator defined in Eq.\ (\ref{Projector}). It selects only the spin-up channels whose transmission probability is characterized by finite values of $\lambda_n$. If these values of $\lambda_n$ are still relatively large (tunneling limit), we can further expand the logarithm to the linear order in small $e^{-2\lambda_n}$. Using the fact that $Q_N$ and $P$ commute we finally obtain the action (\ref{SGamma}). The prefactor is nothing but the normal tunneling conductance according to the Landauer formula \cite{BeenakkerReview}.

\section{\label{General} General solution for Andreev conductance}

In this Appendix we minimize the action (\ref{S1}) with the matrix $Q$ given by Eq.\ (\ref{QthetaW}) and derive the general result for Andreev conductance (\ref{general}). Expansion of the action up to the second order in $W$ and $\chi$ yields
\begin{equation}
 S = S_0 + S_1 + S_2,
\end{equation}
where $S_0$ is given by Eq.\ (\ref{Stheta}). Linear and quadratic terms in the action are
\begin{align}
 S_1
  &= -\frac{G_S}{16} \int_{-\infty}^0 dt \operatorname{Tr} \bigl[
      \sigma_z \tau_y (\dot\theta \dot W + W \sin\theta)
    \bigr] \notag \\
    &\qquad + \frac{G_T}{16} \cos\theta_0 \operatorname{Tr} \bigl[ \sigma_z\tau_y W(0) \bigr], \\
 S_2
  &= \frac{G_S}{32} \int_{-\infty}^0 dt \operatorname{Tr} \biggl[
      \dot W^2 - \frac{\dot\theta^2}{4} \{\sigma_z \tau_z, W \}^2 \notag \\
      &+ W^2 \cos\theta + \frac{3W^2 - (W \mathbf{s})^2}{2 \Delta \tau_\text{SO}}
    \biggr]
    + \frac{G_T}{16} \operatorname{Tr} \biggl[
      \frac{\chi^2}{2} \sin\theta_0 \notag \\
      &+ \left( \frac{W(0)^2}{2} - \chi W(0) \sigma_y \tau_z \right)(\sin\theta_0 + \sigma_z \tau_x s_z)
    \biggr].
 \label{S2}
\end{align}
We observe that the linear part $S_1$ vanishes provided $\theta$ obeys the Usadel equation (\ref{thetaUsadel}) in the bulk and its boundary value is $\theta_0$ from Eq.\ (\ref{theta0}). This is an expected behaviour because the Usadel equation describes a minimum of the action and hence there should be no linear corrections in the vicinity of its solution.

Let us now analyse the quadratic part of the action $S_2$. In the absence of source term $\chi = 0$, this part of the action is minimized by $W = 0$. For a small but nonzero $\chi$, we will look for $W$ linear in $\chi$. The structure of the coupling term in the last line of Eq.\ (\ref{S2}) suggests that only two matrix components are induced in $W$ at the boundary with the ferromagnet:
\begin{equation}
 W
  = \sigma_y \tau_z w_s + \sigma_x \tau_y s_z w_t.
 \label{W}
\end{equation}
Below we will see that this structure with only two components is also preserved in the bulk of the superconductor.

Let us note that parametrization (\ref{W}) looks similar to Eq.\ (\ref{Wweak}) but has an important difference. Both forms of $W$ are linear combinations of a singlet and a triplet component. However, in Eq.\ (\ref{W}) a different structure of the singlet matrix appears as compared to Eq.\ (\ref{Wweak}). This discrepancy has occurred because we are now using the parametrization (\ref{QthetaW}) and expand near different point on the $Q$-matrix manifold compared to Eq.\ (\ref{QW}). The role of the singlet component from Eq.\ (\ref{Wweak}) is now played by the angle $\theta$ in Eq.\ (\ref{QthetaW}).

Substituting Eq.\ (\ref{W}) into Eq.\ (\ref{S2}), we obtain the following bulk and boundary parts of the quadratic action:
\begin{equation}
 S_2
  = S_{2S} + S_{2T},
\end{equation}
\begin{align}
 S_{2S}
  &= \frac{G_S}{4} \int_{-\infty}^0 dt \biggl[
      \dot w_s^2 + \dot w_t^2 - \dot\theta^2 w_t^2 \notag \\
      &\hspace{2cm} + (w_s^2 + w_t^2) \cos\theta + \frac{2 w_t^2}{\Delta \tau_\text{SO}}
    \biggr], \label{S2S} \\
 S_{2T}
  &= \frac{G_T}{4} \biggl[
      \sin\theta_0 \Bigl( \bigl[ \chi - w_s(0) \bigr]^2 + w_t^2(0) \Bigr) \notag \\
      &\hspace{2cm} + 2 \bigl[ \chi - w_s(0) \bigr] w_t(0)
    \biggr]. \label{S2T}
\end{align}
Variation of the bulk part of the action $S_{2S}$ yields equations for $w_{s,t}(t)$ [here we have used Eq.\ (\ref{theta})]:
\begin{align}
 &-\ddot w_s - \frac{2 w_s}{\cosh^2(t - t_0)} + w_s
  = 0, \\
 &-\ddot w_t - \frac{6 w_t}{\cosh^2(t - t_0)} + \left(
   1 + \frac{2}{\Delta\tau_\text{SO}}
 \right) w_t = 0.
\end{align}
Relevant solutions to these equations should decay in the limit $t \to -\infty$ and have the form
\begin{align}
 w_s
  &= \frac{1}{\cosh(t - t_0)}, \\
 w_t
  &= e^{\sqrt{1 + 2/\Delta\tau_\text{SO}}\, t} \biggl[
      \frac{2}{3\Delta\tau_\text{SO}} \hspace{4cm} \notag\\
      &- \sqrt{1 + \frac{2}{\Delta\tau_\text{SO}}} \tanh(t - t_0) + \tanh^2(t - t_0)
    \biggr].
\end{align}

Upon substitution of these functions into Eq.\ (\ref{S2S}), we can reduce the bulk action to an additional boundary term integrating by parts:
\begin{multline}
 S_{2S}
  = \frac{G_S}{4} \Bigl[
      w_s(0) \dot w_s(0) + w_t(0) \dot w_t(0)
    \Bigr] \\
  = \frac{G_S}{4} \Biggl[
      \cos(\theta_0/2) w_s^2(0)
      + \frac{w_t^2(0)}{\cos(\theta_0/2)} \Biggl(
          \cos\theta_0 +
          \Biggl[
            \frac{3}{2}\Delta\tau_\text{SO} \\
            + \frac{2 - 3\Delta\tau_\text{SO}}
              {3 + 2 \sqrt{1 + 2/\Delta\tau_\text{SO}} \cos(\theta_0/2) + \cos\theta_0}
          \Biggr]^{-1}
        \Biggr)
    \Biggr].
\end{multline}
We now add the boundary action $S_{2T}$ from Eq.\ (\ref{S2T}) and obtain the total action as a quadratic expression in $w_{s,t}(0)$ and $\chi$. This total action is to be minimized with respect to $w_{s,t}(0)$ to produce $S_\text{min}(\chi)$. Finally, taking the second derivative in $\chi$ according to Eq. (\ref{GA}) we arrive at the result (\ref{general}).

\bibliography{Andreev2023}

\end{document}